\documentclass[a4paper]{spie}  
\usepackage[]{graphicx}

\title{ABORAS: polarimetric, 10cm/s RV observations of the Sun as a star}

\author{Casper Farret Jentink\supit{a}, Annelies Mortier\supit{c}, Frans Snik\supit{d}, Patrick Dorval\supit{d}, Samantha J. Thompson\supit{c}, Ramon Navarro\supit{b}, and Tim Naylor\supit{e}
\skiplinehalf
\supit{a}Observatoire Astronomique de l’Université de Genève, Chemin Pegasi 51b, 1290 Versoix, Switzerland; \\
\supit{b}NOVA Optical Infrared Instrumentation Group at ASTRON, PO Box 2, Dwingeloo, The Netherlands;\\
\supit{c} University of Cambridge, Astrophysics Group, Cavendish Laboratory, J J Thomson Avenue, Cambridge, UK;\\
\supit{d} Sterrewacht Leiden, Universiteit Leiden, Niels Bohrweg 2, 2333 CA, Leiden, the Netherlands;\\
\supit{e} University of Exeter, Devon, Exeter, EX4 4QL, UK}

\authorinfo{Further author information: (Send correspondence to C.F.J.)\\C.F.J.: E-mail: casper.farret@unige.ch, Telephone: +41 (0)2 23 79 22 92\\A.M.: E-mail: angm2@cam.ac.uk, Telephone: +44 (0)1 22 33 37 299}

\begin{document} 
  \maketitle 

\begin{abstract}
We present a description of \textit{A dual-Beam pOlarimetric Robotic Aperture for the Sun} (ABORAS), to serve as a Solar input with a dedicated Stokes \textit{V} polarimeter for the HARPS3 high-resolution spectrograph. ABORAS has three main science drivers: trying to understand the physics behind stellar variability, tracking the long term stability of HARPS3, and serve as a benchmark for Earth-sized exoplanet detection with HARPS3 by injecting an Earth RV signal into the data. By design, ABORAS will (together with the HARPS3 instrument) be able to measure 10cm/s variations in RV of the integrated Solar disk and detect integrated magnetic field levels at sub 1 Gauss level through circularly polarized light.
\end{abstract}



\section{INTRODUCTION} \label{sec:intro}  
Detecting and characterising small exoplanets with radial velocities (RVs) is currently more hindered by effects arising from stellar variability\cite{robertson2014stellar, davis2017insights,chaplin2019filtering} than by instrumental limitations. Studying the behaviour and origin of these stellar variability signals in RV time series of stars other than our own Sun is inherently difficult since it is incredibly difficult to understand their surfaces. Furthermore, the signal is very likely mixed with planetary induced signals. Given that the typical rms of the Solar RV signal over time lies at a few m/s and an Earth-like planet orbiting a Sun-like star would induce an RV signal of $\sim$10cm/s, Sun-like stars can hide potential exoplanet candidates. Additionally, the RV variations induced by the star can trigger false detections \cite{meunier2010reconstructing}. As such, it is required we gain a better understanding of stellar variability as we try to detect smaller, Earth-mass exoplanets. 

Ongoing efforts have proposed to observe the Sun-as-a-star as a template for RV variability given we can easily resolve its surface with other instruments and it is uncontaminated by unknown planetary signals as we can remove these known effects\footnote{Here we are disregarding Planet Nine as it would create an immeasurably small signal. We estimate the semi-amplitude to be $\sim2$cm/s, but over a period of thousands of years \cite{brown2021orbit}.}. Resolving the Sun allows for spatial cross comparing of unresolved RV data with stellar activity and oscillations. As such, small Solar telescopes that feed into high-resolution (HiRes) stable spectrographs have been commissioned in the past to perform unresolved Sun-as-a-star observations. Examples are LOST at EXPRES (Lowell Observatory) \cite{llama2020lowell}, HELIOS at HARPS, the HARPS-N Solar telescope \cite{HARPS-N_SUN}, and the NEID solar feed \cite{lin2021observing}. They all scramble the image of the Sun, allowing for an RV measurement of the integrated solar disk, treating it effectively as a star. Some designs contain a separate imager to resolve the Sun simultaneously.

Several mitigation strategies have been put forward to deal with stellar-induced RV signals. Some data-driven techniques have been proposed, such as using Gaussian Processes to model the stellar variability as correlated noise\cite{rajpaul2015GPs} or using PCA on the common line profiles\cite{cameron2021Scalpels}. In general, several techniques rely on information from activity indicators. Examples of such indicators are the Full-Width-at-Half-Maximum (FWHM) of the average line profile, the asymmetry of the spectral lines, or chromospheric indicators such as the $\log R'_{HK}$ \cite{figueira2013line, da2011long}. Recently, SDO observations of the Sun have shown that the strongest correlation with activity-driven RVs can be found for hemispherically-averaged unsigned magnetic flux density of the Sun\cite{haywood2016sun,haywood2020unsigned}. However, these observations rely on only one iron absorption line rather than a full optical spectral observation which is how RVs for other stars are extracted\cite{lienhard2022rvextraction}.

In this paper, we propose a possible design for a polarimetric Solar Telescope to be fed into the HARPS3 spectrograph, named ABORAS. It is named after Abora, the (Solar) deity of the Indigenous inhabitants of the Canary Islands. ABORAS is designed to be different from aforementioned Solar telescopes connected to HiRes spectrographs. It will contain a polarimeter to split light into two perpendicularly polarized channels. Aided by the polarimeter, it can measure the amount of circular polarization through Stokes \textit{V}. Light emitted by the Sun is partially circularly polarized due to the Zeeman effect - line splitting that occurs in presence of (strong) magnetic fields \cite{zeeman1897influence}. By measuring the intensity of this effect, we will be able to reconstruct fluctuations in signed magnetic flux from the integrated Solar disk and measured from the full optical spectrum. This will provide more extensive data than previously gathered photospheric full-disk data from for example the SOLIS Vector Spectromagnetograph (VSM), or SDO, which derive magnetic activity from a single line only \cite{bertello2015solis}. We can couple the magnetic information to the directly observed Sun-as-a-star RV fluctuations, measured by the same or similar instrumentation. Currently, the combination of Solar spectropolarimetry coupled with highly stable RVs does not exist, which makes ABORAS a unique and valuable addition to the ensemble of Sun-as-a-star-studying instruments.

\section{The HARPS3 instrument} \label{sec:HARPS3}
HARPS3 (High Accuracy Radial velocity Planet Searcher 3) is a HiRes, echelle type spectrograph that is currently being assembled and aims to start observations in 2024\cite{thompson2016harps3}. HARPS3 is nearly identical in design to HARPS at the ESO 3.6 metre telescope at La Silla, and HARPS-N at the Telescopio Nazionale Galileo (TNG) on La Palma \cite{pepe2002harps,cosentino2012harps}.
The HARPS3 instrument is aiming for a 10cm/s wavelength stability over a 10 year span of observations in the Terra Hunting Experiment (THE). If this criterion is met, it will be able to detect the RV signals exhibited by Earth-sized planets in long-period orbits around Solar-type stars \cite{hall2018feasibility}. The main breakthrough in detection capabilities of HARPS3 in comparison to the past HARPS instruments or other high-precision, high-stability spectrographs is its observing strategy: it will observe a group of stars every night - allowing for long-term as well as short-term RV tracking. A prerequisite for these detections is our capability to mitigate stellar variability to the extent that we can distinguish an exoplanet-induced RV-signal from a star-induced RV signal \cite{hall2018feasibility}.

HARPS3 will be fibre-connected to its own Cassegrain unit (CASS) at the Isaac Newton Telescope (INT) on the island of La Palma. As part of the THE project, the INT will be fully refurbished and roboticized \cite{thompson2016harps3}. Besides improving on wavelength stability and precision, the instrument will house a dedicated polarimeter inside the CASS. The design of the polarimeter has been inspired by the dual-beam exchange polarimeter at HARPS, HARPSpol, which allows for full Stokes characterization and stellar magnetic flux measurements \cite{snik2010harps, snik2013astronomical}.

\begin{table}[ht]
\caption{Summarised science requirements of ABORAS.} 
\label{tab:reqs}
\begin{center}       
\begin{tabular}{|l|l|} 
\hline
\rule[-1ex]{0pt}{3.5ex}  \textbf{Integration time and SNR} & 5 minutes, SNR of $>$300 at 5500 \AA\ and of $>$100 at 4000 \AA.  \\
\hline
\rule[-1ex]{0pt}{3.5ex}  \textbf{Wavelength range} & 380-690 nm   \\
\hline
\rule[-1ex]{0pt}{3.5ex}  \textbf{Polarimetric modes} & Circular ($<$1 Gauss) and depolarized (by continuous QWP rotation).   \\
\hline
\rule[-1ex]{0pt}{3.5ex}  \textbf{Calibrations} & One prior and one post continuous solar observations.   \\
\hline
\rule[-1ex]{0pt}{3.5ex}  \textbf{RV stability and precision} & Identical to HARPS3.   \\
\hline
\rule[-1ex]{0pt}{3.5ex}  \textbf{Solar disk uniformity} & Non-uniformities $<$ 0.01\% (to provide RV stability).  \\
\hline
\rule[-1ex]{0pt}{3.5ex}  \textbf{Visibility} & Airmass $<$3 (19.5 degrees).  \\
\hline
\rule[-1ex]{0pt}{3.5ex}  \textbf{Guiding} & Guiding errors allow for 100\% of the Solar disk to be injected into fibres.  \\
\hline
\rule[-1ex]{0pt}{3.5ex}  \textbf{Cloud/weather detection} & Weather station and solar disk imager for cloud coverage detection. \\
\hline
\rule[-1ex]{0pt}{3.5ex}  \textbf{Data processing} & Provide separate barycentric correction. Use the HARPS3 pipeline.  \\
\hline
\end{tabular}
\end{center}
\end{table}

\section{ABORAS DESIGN} \label{sec:design}

The design of ABORAS was informed by the science requirements that are summarized in Table~\ref{tab:reqs}. These requirements guided us in defining performance, functional, polarimetric, software and interface requirements and ultimately lead us to create the full design. Five minute exposure times are required to reduce the effect of stellar oscillations \cite{chaplin2019filtering}. A SNR of 300 is the recommended SNR to ensure that spectral line asymmetries can be well studied \cite{cretignier2022stellar}. The signed magnetic flux of a quiet solar-type star varies within 10G, so we require to measure better than 1G in order to study the variations. A polarimetric sensitivity of 1G implies a sensitivity of 1E-4 in Stokes V/I \cite{snik2013astronomical}.

\begin{figure}[h]
    \centering
    \includegraphics[width=0.5\textwidth]{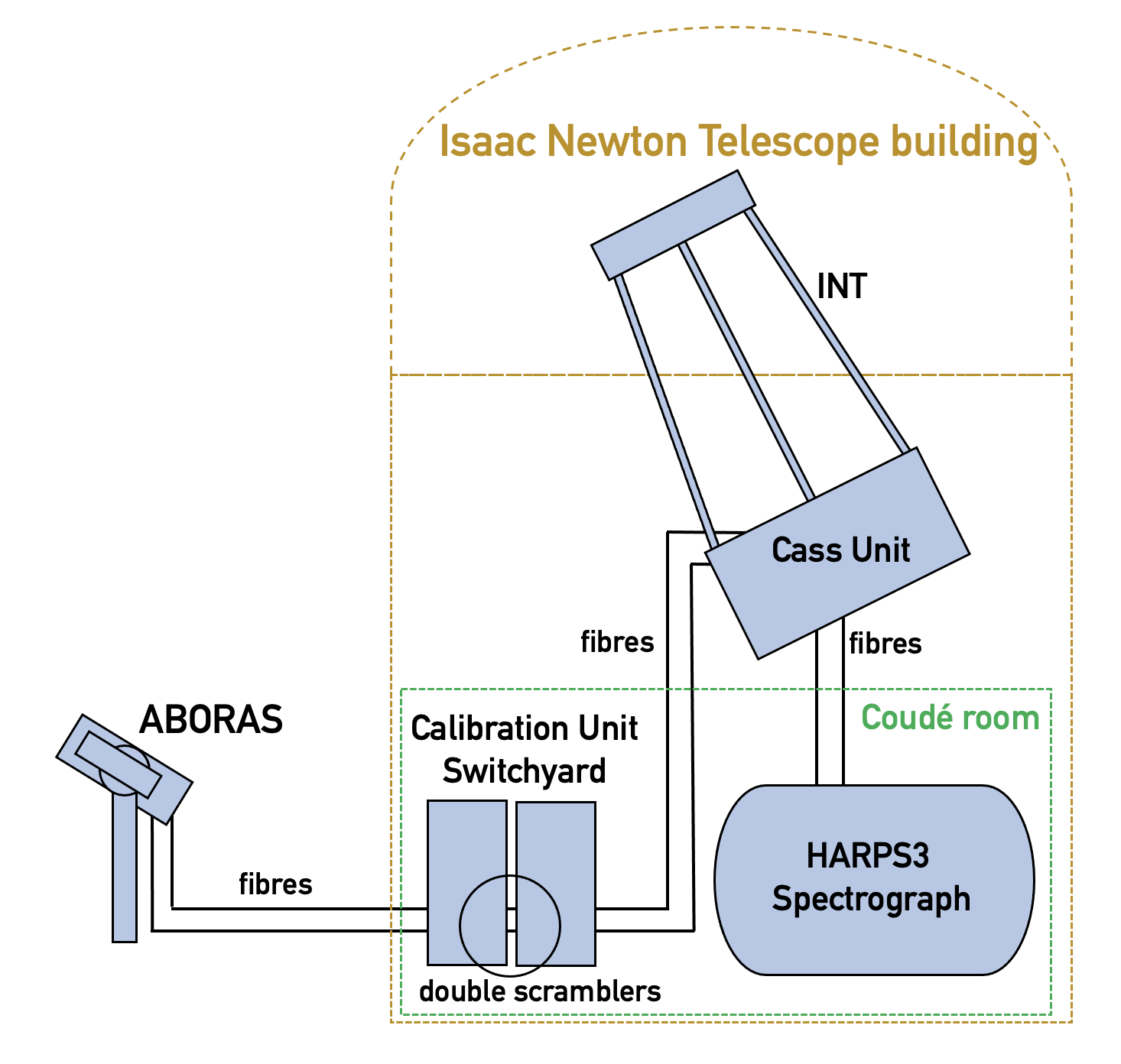}
    \caption{Placement of ABORAS with respect to the Isaac Newton Telescope and major HARPS3 components.}
    \label{fig:placement}
\end{figure}

ABORAS will be fibre-fed to the HARPS3 spectrograph, in a setup displayed in Figure~\ref{fig:placement}. ABORAS has two fibre outputs, each transmitting one perpendicularly polarized component. The HARPS3 instrument contains a calibration unit, which houses a calibration unit switchyard (CUS). The CUS consists of a range of fibre scramblers on a movable rail, allowing observers to change calibration input sources for the HARPS3 spectrograph. We will attach the two fibre outputs of ABORAS to two empty slots on the CUS. To perform Solar observations, the CUS will be aligned to the two fibre inputs that lead to the CASS of HARPS3. Light from the CUS does not travel directly to the spectrograph, to ensure calibration light (or solar light) travels through the same optical path as starlight. Using the CUS to connect ABORAS to the HARPS3 spectrograph allows for a simple connection, without making modifications to the main HARPS3 design.

ABORAS itself consists of a 4mm pinhole type entrance aperture, a dual-beam exchange method polarimeter, and two integrating spheres, allowing us to observe the Sun as a star by integrating the full solar disk. From the integrating spheres, optical fibres lead to the CUS. All of the optics are enclosed in a weather-proof housing, mounted to a commercial solar tracking system which uses a pyrheliometer to track the Sun to specification.

\subsection{Dual-beam exchange method polarimeter}
We have optimised the design of ABORAS to uniformly capture circularly polarized sunlight, split along perpendicular polarization states. The polarimeter of ABORAS is of a dual-beam exchange type, allowing for a high polarimetric sensitivity due to simultaneous perpendicular polarization state feeds. The polarimeter consists of a rotating superachromatic quarter waveplate (QWP) following a Pancharatnam design \cite{samoylov2004achromatic}, and a polarizing wire-grid beamsplitter. This setup enables us to measure the amount of circular polarization (Stokes \textit{V/I}) through multiple spectral intensity recordings. We decided not to include a half-waveplate, which would have added capabilities for determining the linear Stokes parameters, \textit{Q}, and \textit{U}. We will only be able to detect circular polarization given its first order dependence on the magnetic field strength. The second-order dependence for linear polarization states is a result from a linearly polarized, non-splitted component induced by the Zeeman effect \cite{crutcher2019review}.

To limit the amount of crosstalk induced by the optics in front of the polarimeter, we limit ourselves to a single longpass filter with a cutoff at 380nm since HARPS3 is not designed to work below this wavelength anyway. This filter acts as an optical window and prevents UV flux from damaging the QWP. Since we opt for a QWP based on 5 layers of zero-order birefringent polymer, we expect UV light to cause photo-oxidation of these polymer layers \cite{feldman2002polymer}. In order to minimize heat load, we chose a 4mm diameter pinhole as entrance aperture, still ensuring enough flux to reach the spectrograph, despite the integrating spheres.

\subsection{Polarimetric accuracy and efficiency}
\begin{figure}
   \begin{center}
   \begin{tabular}{cc}
   \includegraphics[height=5cm]{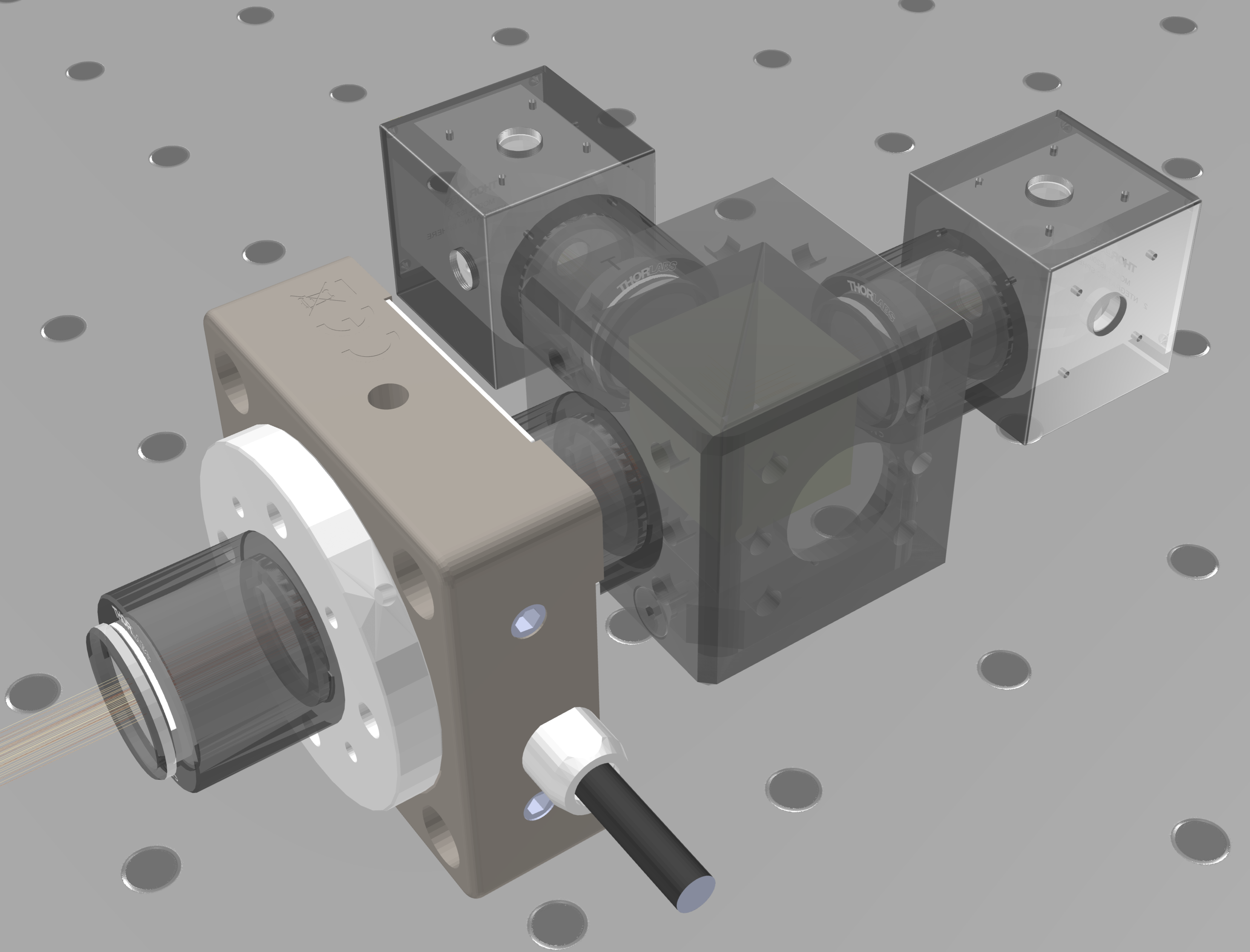}&\includegraphics[height=5cm]{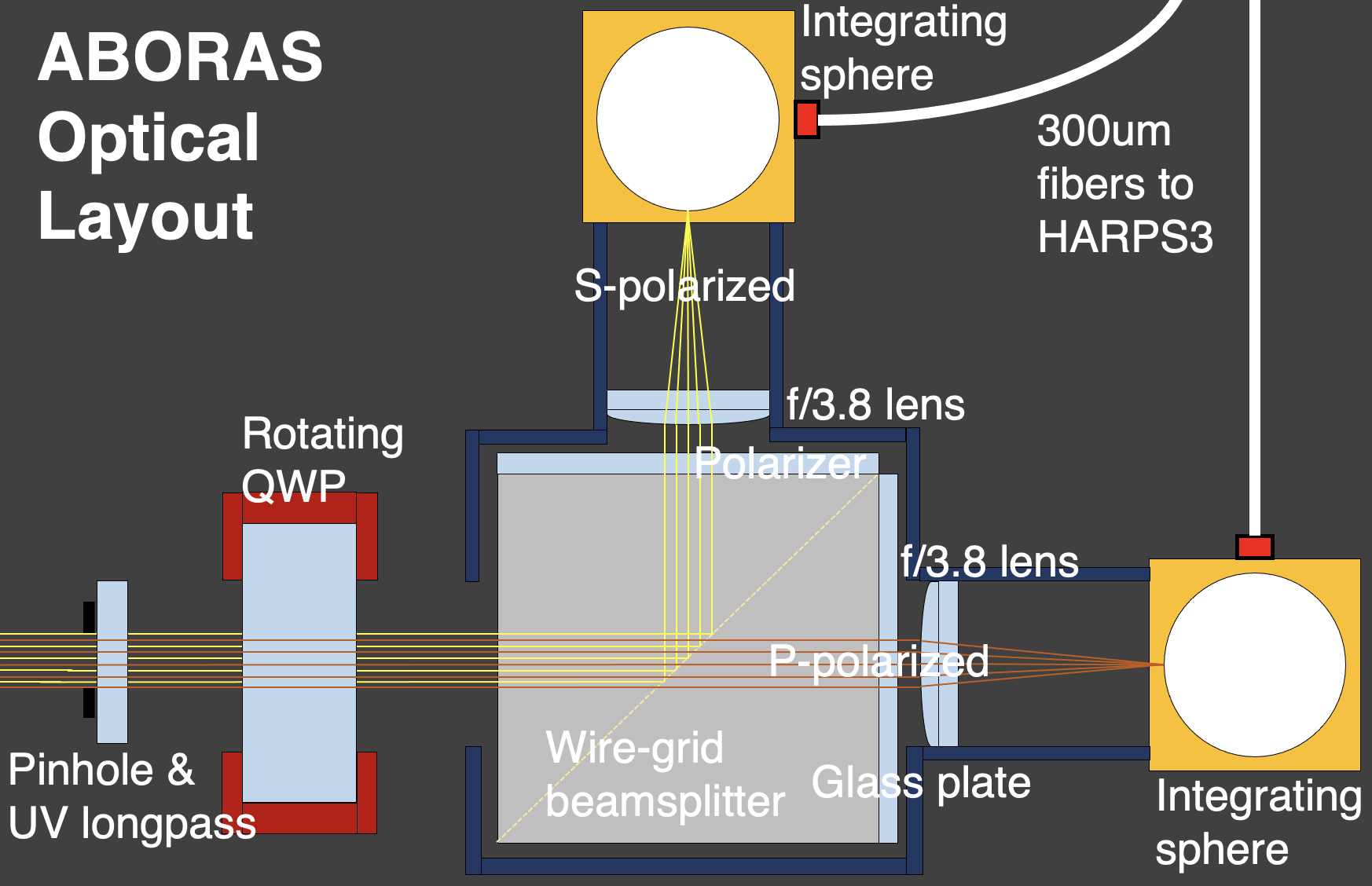}\\
   \end{tabular}
   \end{center}
   \caption[example] 
   { \label{fig:design} 
On the left we find a 3D optical model of ABORAS. From left to right we find the UV longpass filter, a Piezo driven rotation stage containing the super achromatic quarter waveplate, the polarizing wire-grid beamsplitter with a linear polarizer at the reflected beam and a glass plate at the transmitted beam, and two f=19.0mm achromatic doublets focussing the light into 10mm integrating spheres. In the right figure we find a labeled sketch of the optical train.}
   \end{figure}

The polarimetric performance requirements of ABORAS were adopted from the polarimetric requirements of the CASS. The relevant polarimetric requirements state: The scaling of Stokes \textit{V} is accurate $> 95\%$, either by calibration or design; The polarimetric efficiency is $> 90\%$ over the entire wavelength range (380-690nm). We decided to match polarimetric requirements of the HARPS3 polarimeter and ABORAS as to ensure polarimetric measurements of the Sun and target stars of the THE are comparable. As a consequence, the QWP, wire-grid polarizing beamsplitter, and linear polarizer of ABORAS adopt their design from the CASS.\cite{dorval2018analysis} In the optical path (Fig.~\ref{fig:design}), only the UV longpass filter resides in front of the polarimeter. To estimate how much a longpass filter influences the polarimetric performance of ABORAS, we performed measurements on an in-house, unbranded UV filter with a dual-rotating polarizer measurement set-up.  The exact properties of the tested UV filter, like the sharpness and location of the cutoff wavelength, are unknown. It is important to note that longpass filters come in various variants, of which the two most commonly used are absorption-, and interference-based filters. It will be critical to repeat our measurements for the final filter of ABORAS as we do not know whether all filters show similar behaviour. The measurements and results below thus only suffice as an initial test of the measurement setup and as a check to determine if we can expect a UV filter to live up to design requirements.

\begin{figure}[h]
    \centering
    \includegraphics[width=0.65\textwidth]{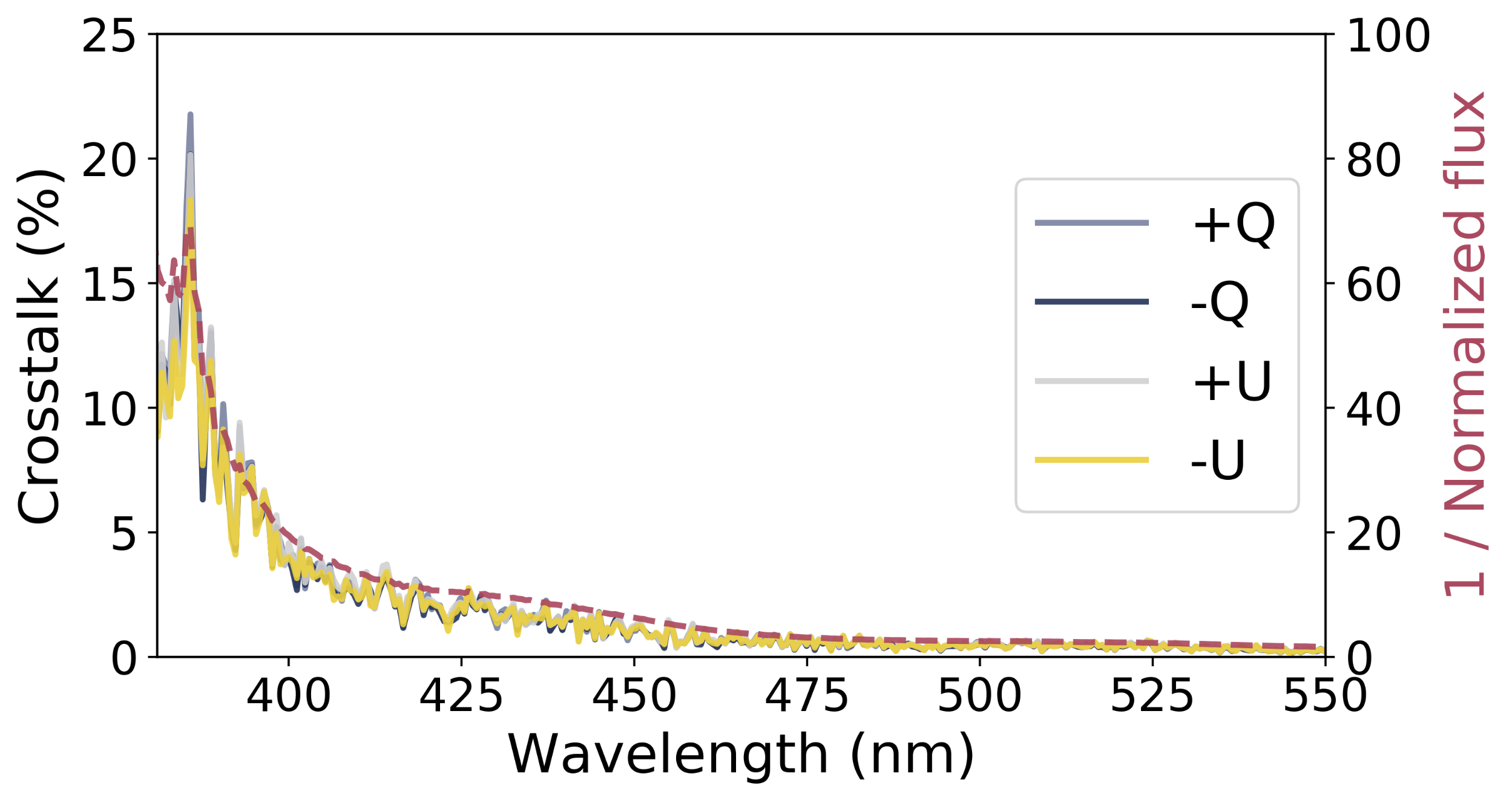}
    \caption{On the horizontal axis we find the wavelength and on the left vertical axis we find the percentage of crosstalk. The colored lines show the upper bounds on the uncertainty margins for the percentage of crosstalk induced by the UV filter for all 4 linear polarization states: $\pm$ \textit{Q/U}. We see that the lines mostly overlap. In the same plot on the right vertical axis we find the inverse of the spectrum of the light source that was used in the measurement setup, denoted with the dashed line. The strong correlation between the inverse of the flux and upper margin for the percentage of crosstalk indicates our measurements are noise limited. Values above 550nm were cut from the plot, as all values are near zero.}
    \label{fig:crosstalk}
\end{figure}

In Figure~\ref{fig:crosstalk} the upper margin on crosstalk for each linear polarization state of the UV filter can be found. The values show the upper limit on retardance - estimated from the offset with a perpendicular polarizer power measurement. In the same plot the inverse of the spectrum of the light source is shown. A strong correlation is seen between the inverse spectrum and maximum threshold on crosstalk for all polarization states. This indicates the measurements were largely noise limited. We were not able to determine whether the UV filter meets requirements on polarimetric efficiency below 400nm. However, given its excellent performance for wavelengths above 550nm, we do not expect the filter to behave very differently at shorter wavelengths closer to the cutoff. Also, for a dielectric interference filter, knowing that the angle of incidence is close to perpendicular, any polarization dependent effects should be negligible.

\subsection{Scrambling and fibre injection}

\begin{figure}[!b]
    \centering
    \includegraphics[width=0.99\textwidth]{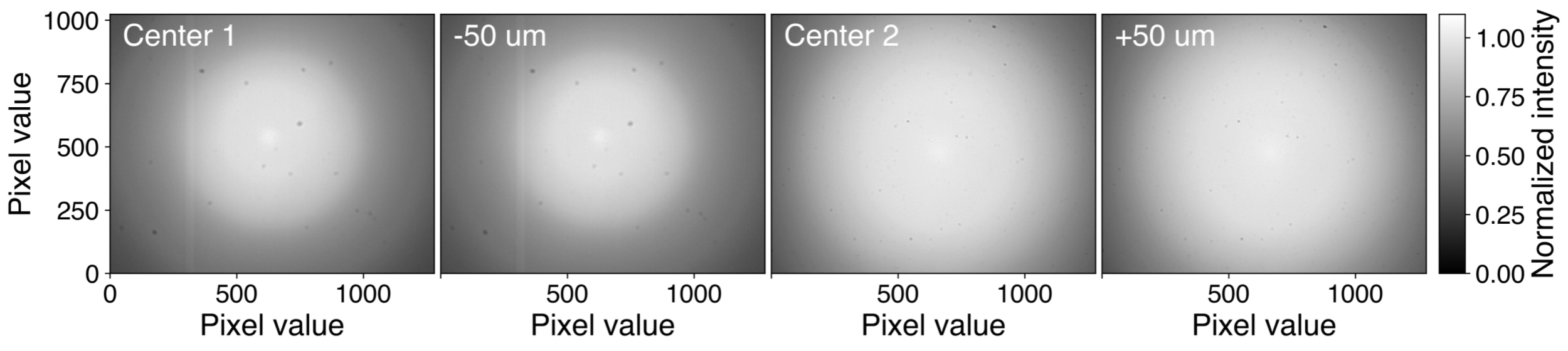}
    \caption{Four far-field measurements at the output of a 50cm long 300$\mu$m circular fibre, connected to the output port of the integrating sphere. Note that for the -50$\mu$m offset measurement we found a method to mount the fibre head closer to the detector. As such we sample a larger region of the far-field. For the Center 1 and -50$\mu$m frame some bad CCD columns can be seen. All measurements contain some dust that was present on the detector. Both the -50$\mu$m and +50$\mu$m measurement have their own center frame (with no offset from the centre). Given the high losses of the sphere, each field is a stack of 3000, 1-second integration time frames, to reach proper signal-to-noise levels. }
    \label{fig:farfields}
\end{figure}

After light has travelled through the polarimeter, the reflected beam passes through a linear polarizer to improve the polarimetric accuracy of the S-polarized beam. The transmitted P-polarized beam (which by design already meets polarimetric accuracy requirements) passes through a glass plate, to ensure an equal optical path length for both beams. The perpendicularly polarized beams are then focused into 10mm integrating spheres by achromatic doublet lenses with focal lengths of 19mm. The integrating spheres are fibre coupled to the HARPS3 calibration unit switchyard, connected to the spectrograph. The integrating spheres scramble the image of the Solar disk to a uniform response over the Solar disk up to the $10^{-4}$ level, ensuring $<$10cm/s stability in radial velocity precision of the integrated Solar disk.  

\begin{figure}[!h]
   \begin{center}
   \begin{tabular}{cc}
   \includegraphics[height=6cm]{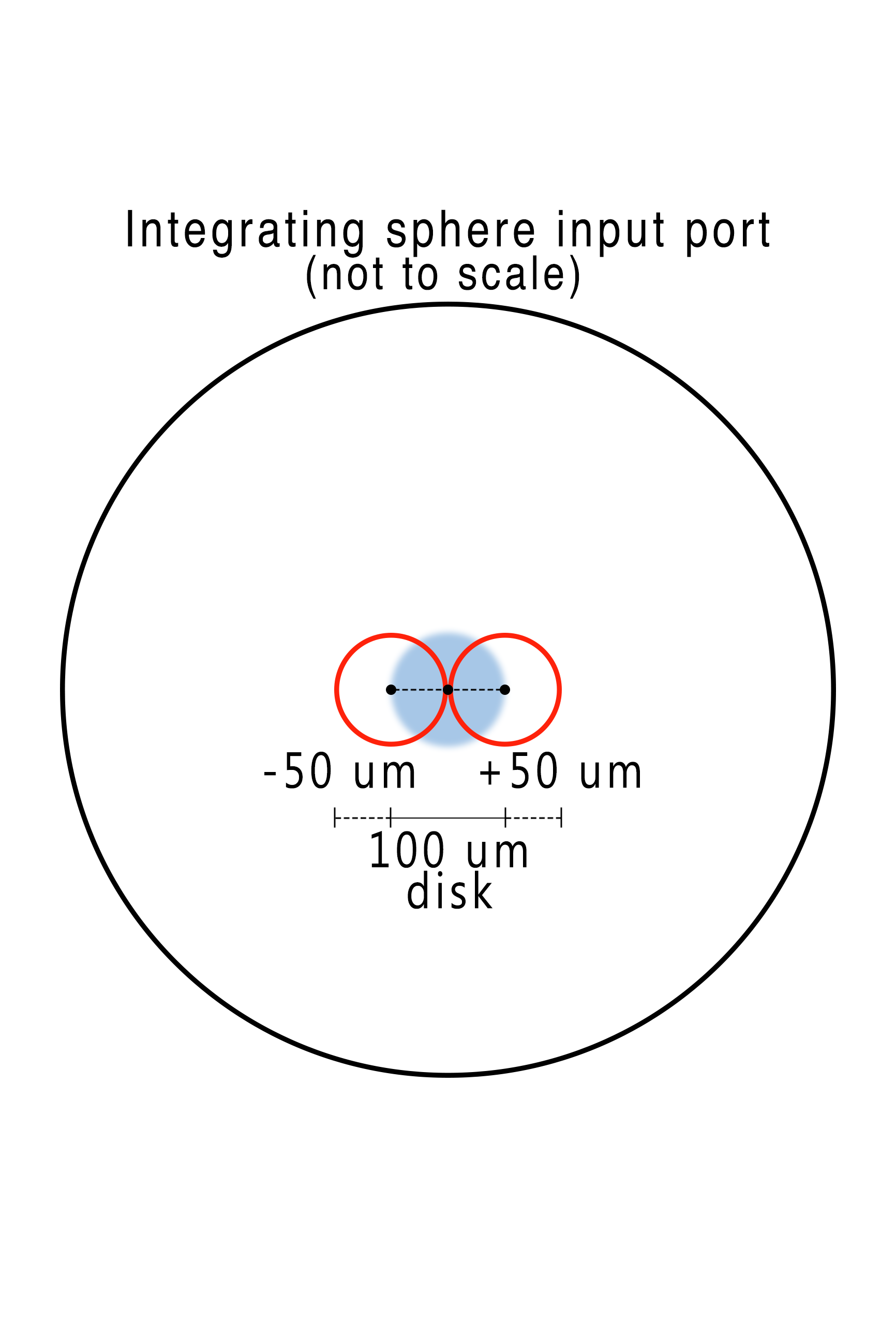}&\includegraphics[height=5cm]{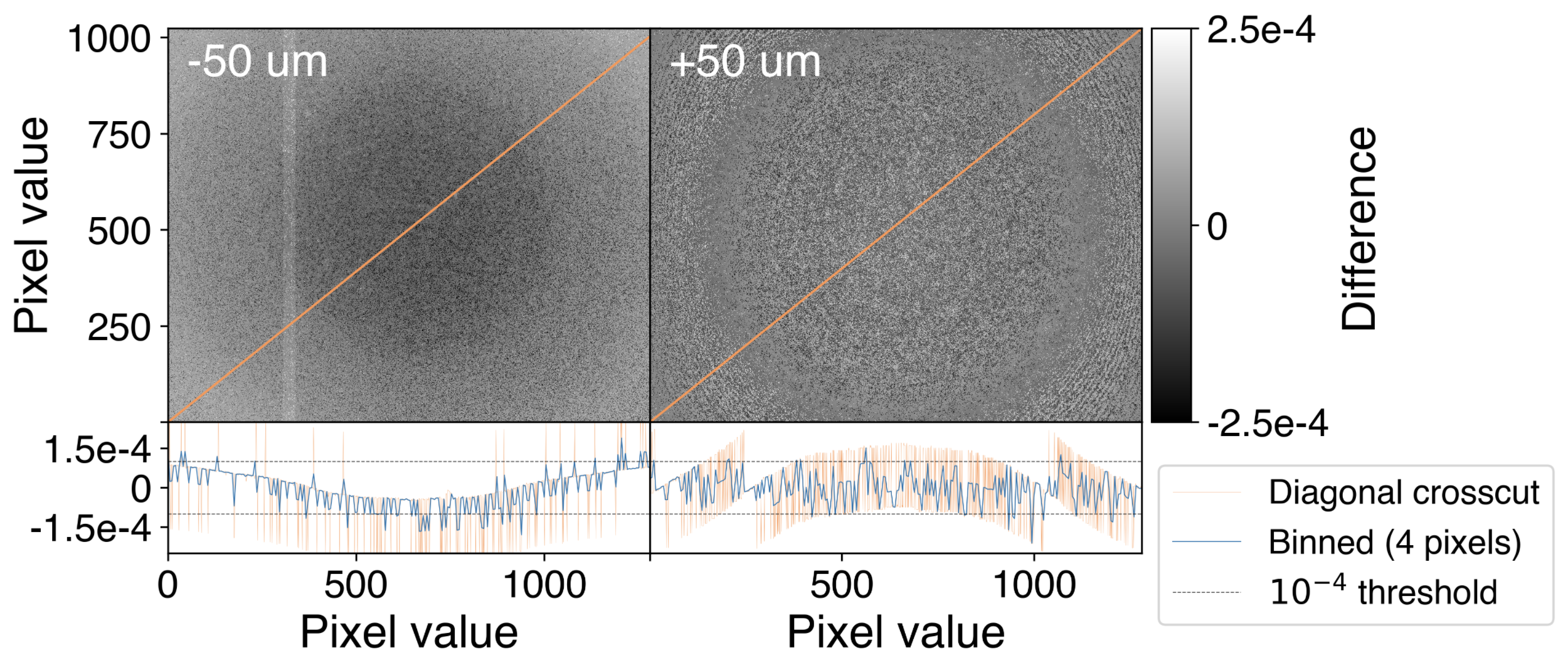}\\
   \end{tabular}
   \end{center}
   \caption[example] 
   { \label{fig:Intsphere} 
The integrating sphere scrambling efficiency tests were performed by injecting a 100$\mu$m disk of light emitted by a super-continuum laser at the center of the input port. This measurement was repeated for a slight offset of 50$\mu$m (left Figure). The far-field was imaged for both measurements and the central and offset frame were subtracted. The residuals show minor structure at the $10^{-4}$ level.}
   \end{figure}

Integrating spheres were chosen to ensure a stable injection of the entire solar disk at high uniformity into the fibres. For the aimed 10cm/s RV precision of the HARPS3 instrument, it is essential that ABORAS is sensitive to the entire Solar disk at a uniformity level of $10^{-4}$ because stellar surface processes happen quasi-randomly across the full Solar surface\cite{dravins2021spatially}. Sunspots at specific locations on the solar surface can for example introduce systematic offsets in RV by blocking part of the light. Given fibre head irregularities and potential dust accumulation throughout the lifespan of the instrument, we dismissed the option of fibre injection by means of focussing the image of the Solar disk on the fibre heads. A static, highly stable solution was found in integrating spheres. For the ABORAS design, we chose small, 10mm spheres as to limit flux loss, since the relation between losses and sphere radius scales by the area of the reflective surface ($R^{3}$)\cite{fussell1974approximate}. While an integrating sphere for Solar disk injection has been used in aforementioned solar telescopes and found to work well \cite{llama2020lowell,HARPS-N_SUN, lin2021observing}, these were significantly larger at 50mm in diameter. To determine the scrambling efficiency of a 10mm integrating sphere, we conducted injection tests for various focal plane positions at the input port of a sphere. For our experiments we tested the polymer P10 integrating sphere manufactured by Artifex Engineering. We injected an artificial 100$\mu$m solar disk at the center of the input port and imaged the far-field at the output port of the sphere. We repeated this test for two slight offsets of $\pm 50\mu$m. The imaged far-fields can be found in Figure~\ref{fig:farfields}. By looking at the residuals of the two subtracted fields, as seen in Figure~\ref{fig:Intsphere}, the difference between the center frame with respect to the offset frame, we concluded that some minor structure was visible. However, from the diagonal crosscuts, we see that the noise level of the measured values is close to our requirement level. To improve on our signal-to-noise, the diagonal cut was also binned over 4 pixels (on the horizontal axis). We find that for both measurements the residual structure stays at or below our $10^{-4}$ threshold. As such, we are confident that scrambling efficiency meets our requirements for a 100$\mu$m solar disk focal plane injected at the input port, including a significant pointing margin.

\subsection{Integrating sphere losses}
The throughput of the Artifex Engineering P10 was measured at three visible wavelengths, spanning the HARPS3 wavelength band: 406, 520, and 660nm. The corresponding loss factors were found to be $11.95 \cdot 10^3$, $10.05 \cdot 10^3$, and $9.00 \cdot 10^3$. The theoretical loss factor that follows from the integrating sphere radiance equation \cite{fussell1974approximate} is in good agreement with our results in combination with the 0.22NA 300$\mu$m fibre connected to the output port. We note that the tested integrating sphere contains two output ports of 3mm in diameter, whereas ABORAS requires only one. We estimate that by reducing the diameter of the input (also 3mm) and output port by a factor of 2, we can reach a loss factor close to $2.4 \cdot 10^3$, which satisfies our requirements.

\subsection{Expected integration times and SNR}
Based on the aperture size of 4mm, in combination with the expected losses of the polarimeter, 10mm integrating spheres adapted to aforementioned modifications, and fibres, we expect integration times of 5 minutes to easily reach an SNR of 300 at 5500 \AA\ on the HARPS3 spectrograph, fulfilling our science requirements. Values were derived by scaling our expected throughput to the HARPS-N Solar telescope design and its typical integration times to reach the same SNR ($<$30 seconds), given the closely identical design of HARPS3 and HARPS-N. To make this comparison, we use an entrance aperture 60 times smaller in size, a 50\% loss at the polarimeter, but 10 times longer integration times. Additionally, we estimate the ABORAS integrating spheres to have a 40 times higher throughput with respect to the HARPS-N Solar telescope integrating sphere, based on the integrating sphere radiance equation \cite{fussell1974approximate}. This easily compensates for our reduced entrance aperture size. We propose to attenuate the flux from the solar telescope with the variable continuous neutral density (ND) filter (OD0 to OD4) in the HARPS3 calibration unit switchyard, to reach desired levels. The ND filter is present in HARPS3 to simulate various magnitudes for observed stars, mitigating the effect of short-period solar oscillations like p-modes. We estimate the range of possible integration times with this filter at an SNR of 300 at 5500 \AA\ to be 180 seconds to a full day. However, it is important to note that the latter is just an extrapolation of the ND filter range, which does not account for detector dark current and potential RV stability issues.

\subsection{Enclosure, pointing, guiding, and imager}
Given the polarimetric capabilities of ABORAS, the telescope cannot be obstructed by any angled glass or transparent plastic, especially if it is curved. Convenient, static, transparent solutions like an acrylic dome, as used for the HARPS and HARPS-N solar telescope \cite{HARPS-N_SUN}, would destroy any polarimetric signal through stress birefringence. A solution to our issue was found in a commercial active solar tracking (AST) system. A similar setup was used for the NEID solar feed \cite{lin2021observing}. Most systems, from various manufacturers, are dust and water tight with IP65 certification, implying the mount does not require an enclosure. Besides, most AST systems provide a tracking accuracy $<1'$ through a pyrheliometer, easily satisfying our science requirements. As a consequence, the optics of ABORAS are to be housed in a dust and watertight enclosure, mounted to the AST system. The UV longpass filter acts as an optical window. Dust accumulation on the optical window is to be prevented with a Sun-cap and regular cleaning. The Sun-cap will also limit scattered light.

For the imager we propose any commercially available CCD based imager to take at least one image of the Sun for every HARPS3 solar exposure. A commercial system is cost efficient and can easily be replaced in case of failure throughout the lifespan of ABORAS. Replacement of the solar imager will not impact any of our performance requirements, as it is a separate system and does not influence the Solar spectra.

\section{Conclusion} 
We have developed a design for ABORAS: a novel polarimetric Sun-as-a-star telescope, which we propose to be fibre-fed into the HARPS3 high-stability, high-resolution spectrograph. ABORAS has been designed around its polarimeter: by carefully adapting design heritage from other Sun-as-a-star telescopes, we were able to include polarimetric capabilities. The dual-beam exchange method polarimeter of ABORAS allows for measurement of Stokes \textit{V/I}, which allows us to reconstruct signed magnetic flux variations on the short, and long term. In combination with highly stable Solar RVs we will search for a relation between RV variations and the disc-integrated magnetic flux density. Ultimately, a better understanding of RV variability on short and long timescales will allow us to unmask Earth-like signals from the stellar induced component paving the way for the detection of a true Earth twin.

The inclusion of a polarimeter in ABORAS required us to re-think some components that have traditionally been used in past, similar telescopes. For ABORAS, we propose to re-scale the integrating sphere(s) to a smaller size, limiting the flux losses, ultimately allowing us to down-scale the entrance aperture size. This was a necessity to limit flux passing through the QWP in the polarimeter. We performed lab tests to confirm our integrating spheres would not induce systematic RV uncertainties higher than 10cm/s due to lower scrambling efficiency. Additionally, we conducted measurements on a UV longpass filter with a dual-rotating polarizer set-up, to test if such a component in front of our polarimeter induces crosstalk and as a result lowers polarimetric accuracy and efficiency. Our measurements were noise-limited, but showed good performance above 400nm. At shorter wavelengths, we expect similar performance but we can recommend a repetition of these measurements for the final filter that is to be fitted to ABORAS.

Overall, we estimate that ABORAS will help us in reaching the 10cm/s threshold in RV precision by mitigating the stellar variability component, necessary to detect Earth-like planets in the habitable zones of their host stars. This will lead to an increased understanding on the occurrence rate of Earth-mass exoplanets and how Earth itself fits into this ensemble of planets.


\acknowledgments     
The authors would like to acknowledge the extensive lab support from collaborators at the NOVA Optical Infrared Instrumentation group in Dwingeloo and Observatory of Geneva. This work was supported by the Swiss National Science Foundation and the Kavli Institute for Cosmology in Cambridge. The authors acknowledge the financial support of the SNSF and KICC.

\section*{SOFTWARE}
\texttt{numpy} \cite{van2011numpy}
\texttt{scipy} \cite{virtanen2020scipy}
\texttt{matplotlib} \cite{hunter2007matplotlib}

\bibliography{report}   
\bibliographystyle{spiebib}   

\end{document}